\documentclass[a4paper,prb,aps,epsfig,twocolumn,showpacs,preprintnumbers,superscriptaddress,float,amsmath,amssymb]{revtex4}

\usepackage{graphicx}
\usepackage{dcolumn}
\usepackage{bm}

\usepackage{graphicx}
\usepackage{pgfplots}
\usetikzlibrary{arrows.meta}
\pgfplotsset{compat=1.16}

\begin{document}

\title{Topology in non-linear mechanical systems}
\date{\today}

\author{Po-Wei Lo}
\affiliation{Laboratory of Atomic and Solid State Physics,
Cornell University, Ithaca, NY, 14853}
\author{Krishanu Roychowdhury}
\affiliation{Department of Physics, Stockholm University, SE-106 91 Stockholm, Sweden}
\author{Bryan Gin-ge Chen}
\affiliation{Lorentz Institute for Theoretical Physics, Leiden University, Leiden 2333 CA, Netherlands}
\author{Christian D Santangelo}
\affiliation{Department of Physics, Syracuse University, Syracuse, NY, 13244}
\author{Chao-Ming Jian}
\affiliation{Laboratory of Atomic and Solid State Physics,
Cornell University, Ithaca, NY, 14853}
\author{Michael J Lawler}
\affiliation{Laboratory of Atomic and Solid State Physics,
Cornell University, Ithaca, NY, 14853}
\affiliation{Department of Physics, Applied Physics and Astronomy, Binghamton University, Binghamton, New York 13902}

\begin{abstract}
Many advancements have been made in the field of topological mechanics. The majority of the works, however, concerns the topological invariant in a linear theory. We, in this work, present a generic prescription of defining topological indices which accommodates non-linear effects in mechanical systems without taking any approximation. Invoking the tools of differential geometry, a $\mathbb{Z}$-valued quantity in terms of the Poincaré–Hopf index, that features the topological invariant of non-linear zero modes (ZMs), is predicted. We further identify one type of topologically protected solitons that are robust to disorders. Our prescription constitutes a new direction of searching for novel topologically protected non-linear ZMs in the future.

\end{abstract}

\maketitle

Mechanical systems offer a remarkable connection between physics and engineering. Through their simplicity, they have inspired both ideas at the foundation of theoretical physics and a sense of control over our physical world. In the recent field of topological condensed matter, following hints that topology can play a role in non-linear fine-tuned mechanical systems\cite{lawler2013emergent}, Kane and Lubensky\cite{tm1} uncovered a connection between topological insulators\cite{ti1,ti2,ti3,ti4} and linearized balls-and-springs models. With importance in the field of metamaterials\cite{tm2,tm3,tm4,tm5,tm6,tm7,tm8,tm9,tm10,tm11,tm12,tm13,tm14,tm15,tm16} and magnetics\cite{origami2,roychowdhury2018classification}, they realized if constraints define the system, zero modes (ZMs) can be topologically protected by TKNN-like topological invariant\cite{qhe}.

It was quickly realized that Kane and Lubensky's ZMs in the case of a chain model they construct can survive back into the non-linear regime and become bulk solitons\cite{sol}. But a formally identical origami system was identified that does not exhibit these solitons\cite{origami1}.
More non-linear ZMs were found in mechanical systems in numerical simulations\cite{2d1,2d2}. In a one-dimensional chain, a domain wall separating two distinct polarizations can be identified by constructing a sequence of consecutive maps on the space of ZMs of a single unit cell\cite{sol2}. However, that does not quite guarantee that this domain wall can move continuously along the chain like a soliton. Thus, the existence of a soliton relies on the exact parameters of a model\cite{sol3}. To the best of our knowledge, however, it remains unclear if solitons observed in generic mechanical systems are always topologically protected or not, and if so, what is the topology to classify them?


In this paper, we develop an exact theory to study the topological invariant for the kinematics of periodic mechanisms satisfying holonomic constraints such as those that arise in {\it e.g.} linkages and origami. 
Using the concept of differential geometry, our theory predicts the existence of a $\mathbb{Z}$-type topological invariant which can then be used to understand which features of the non-linear ZMs of a mechanical system are topologically protected. Applying this to the Kane-Lubensky (KL) chain, we realize the topology to classify the (two) distinct phases of the KL chain, namely the ``flipper'' and the ``spinner'', and further show that the existence of the flipper soliton is topologically protected and robust to disorders (unlike the spinner). In distinction, the origami chain does not support any soliton despite the superficial similarity of its linear ZMs to those of the KL chain.



We start by characterizing the type of mechanical system we are interested in. We assume that the state of the system can be described by generalized degrees of freedom, $\boldsymbol{\theta} = (\theta_1, \theta_2, \cdots, \theta_n)$, and that the system is characterized by a set of (spring) extensions
$\mathbf{e}(\boldsymbol{\theta}) = ( e_1(\boldsymbol{\theta}), \cdots, e_m(\boldsymbol{\theta}) )$. While the elastic energy of such a system can be written as $E(\boldsymbol{\theta}) = \sum_i k_i e_i(\boldsymbol{\theta})^2$ for a set of moduli $k_i > 0$, here we will only be interested in the ground state configurations specified by $\bar{\boldsymbol{\theta}}$ such that $\mathbf{e}({\bar{\boldsymbol{\theta}}}) = \mathbf{0}$. 
If we work with a mechanical linkage or a spring network as in Ref.~\onlinecite{tm1}, we can think of $\boldsymbol{\theta}$ representing the positions of the vertices of our network and $e_i(\boldsymbol{\theta})$, the extension of the springs (from their equilibrium lengths). In this language, the Jacobian $\partial e_i(\boldsymbol{\theta})/\partial \theta_j$ is termed the rigidity matrix.


\begin{figure}
  \includegraphics[width=8cm]{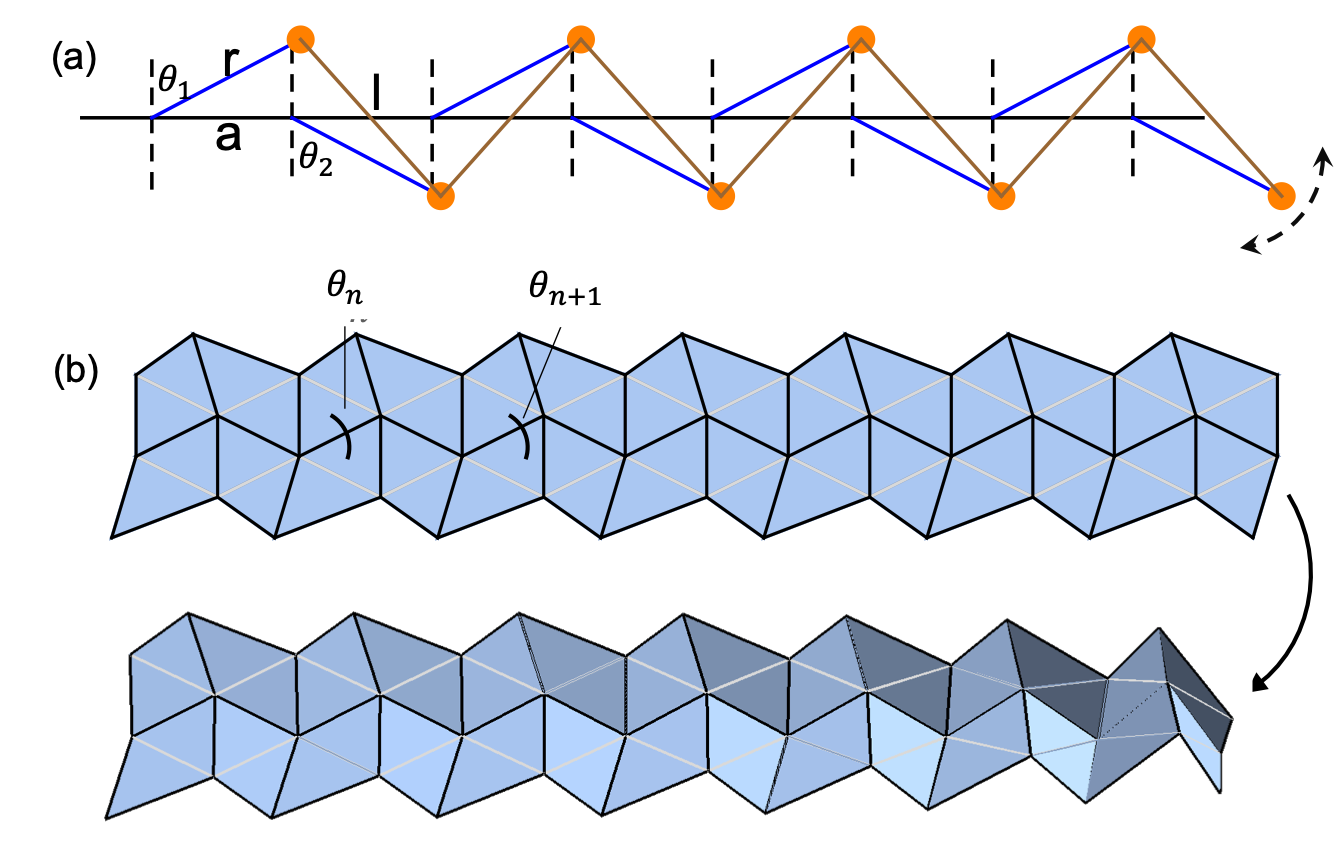}
  \caption{(a) The KL chain has an edge mode on either the left or right edge. (b) The origami chain has an edge mode on either the left or right edge.}
  \label{fig:KLchain}
\end{figure}

Before presenting our prescription of defining topological indices, it is useful to review two examples that pose some apparent paradoxes in defining the topological invariant of the linear ZMs. First, for the KL chain, it is often easier to express the generalized coordinates in terms of the rotation angle of a series of rotors so that $\theta_i$ is the angle between the $i^{th}$ rotor and the vertical axis as shown in Fig. \ref{fig:KLchain} (a). The extension of the $i^{th}$ spring which connects the $i^{th}$ rotor with the $(i+1)^{th}$ rotor then takes the form
$e_i( \boldsymbol{\theta} ) = f(\theta_i, \theta_{i+1}),$
where 
\begin{eqnarray}\label{eq:fourbarconstraint}
    f(\theta_i, \theta_{i+1}) &=&[ (a + r \sin \theta_{i+1} - r \sin \theta_i)^2 \\ \nonumber
    & & + (r \cos \theta_{i+1} + r \cos \theta_i)^2]^{1/2} - L,
\end{eqnarray}
$a$, $r$, and $L$ are the distance between two consecutive pivot points, the radius of the rotors, and the equilibrium length of the springs, respectively. For an open chain of $n$ springs (and $n+1$ rotors), if we choose $\theta_{n+1} = \theta_{1}$, then we have exactly as many constraints as the degrees of freedom, making the system isostatic.

In the second example of the origami chain\cite{origami1}, we instead use $\theta_i$ to denote the supplement of the dihedral angle of one of the folds of each vertex, also called the fold angle [Fig. \ref{fig:KLchain} (b)] (see Appendix \ref{sec:origami}). In this case,
\begin{equation}\label{eq:origamiconstraint}
    f(\theta_i, \theta_{i+1}) = A \sin^2 ( \theta_i/2 ) - B \sin^2 (\theta_{i+1}/2) + \epsilon,
\end{equation}
where $0 < A < 1$, $0 < B < 1$, and $\epsilon$ are defined in Appendix \ref{sec:origami}. While it is straightforward to generalize the above equations to any periodic structure, for simplicity, we specialize to the examples mentioned above focusing on Eq. (\ref{eq:fourbarconstraint})-(\ref{eq:origamiconstraint}) for the remainder of this paper. 

In both the KL chain and the origami chain, if we assume a uniform solution of $\mathbf{e}({\bar{\boldsymbol{\theta}}}) = \mathbf{0}$,
following Ref. \onlinecite{tm1}, the polarization is defined as the integer
\begin{equation}\label{eq:Qdef}
    Q = \frac{1}{2\pi i} \int_{\pi}^\pi {\rm d}q~ \frac{\partial}{\partial q} \ln \left[ \partial_1 f(\bar{\theta},\bar{\theta}) + \partial_2 f(\bar{\theta}, \bar{\theta}) e^{i q} \right].
\end{equation}
where $\partial_a$ implies the derivative with respect to the $a^{th}$ variable in the argument of $f$. When $|\partial_2 f(\bar{\theta},\bar{\theta})| > |\partial_1 f(\bar{\theta},\bar{\theta})|$, $Q = 0$ and when $|\partial_2 f(\bar{\theta},\bar{\theta})| < |\partial_1 f(\bar{\theta},\bar{\theta})|$, $Q = 1$. These two values of $Q$ define two distinct topological phases. For finite systems, the bulk is rigid for both $Q=0$ and $1$, however, the feature that distinguishes these two phases is the location of the linear ZM. 

The behavior above is exhibited by the linear ZMs in both the KL chain and the origami chain, as it should. But in the KL chain (and not the origami chain), certain non-linear deformations can propagate across the system resulting in the edge mode appearing on the other side. In that sense, the polarization defined by Eq. (\ref{eq:Qdef}), though an integer, is not necessarily topologically robust.



\begin{figure}
  \includegraphics[width=8cm]{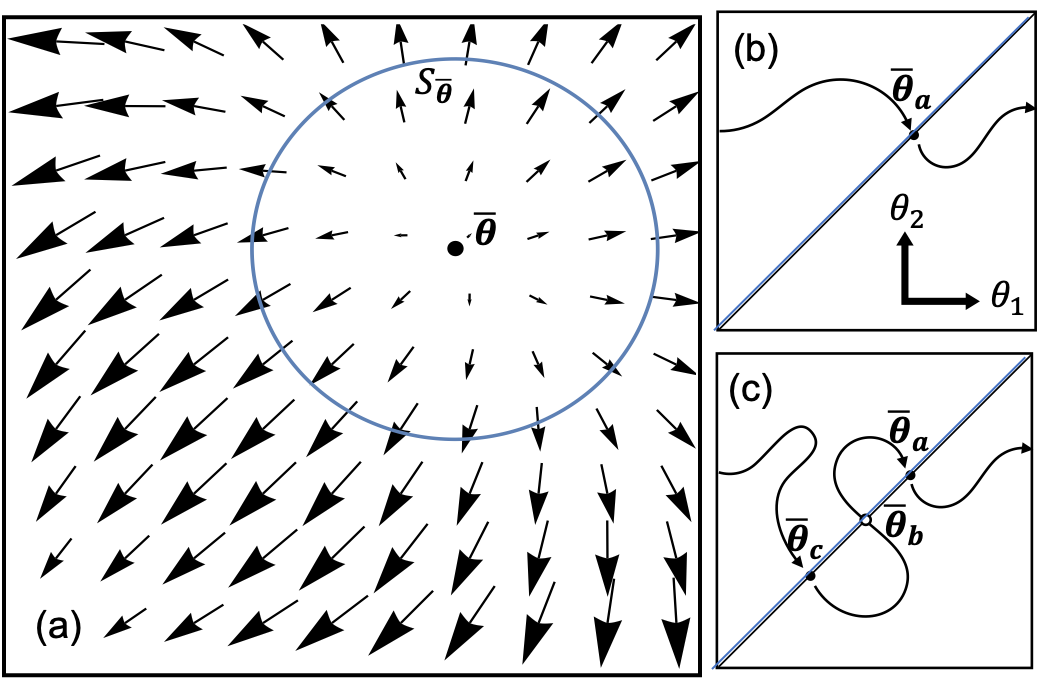}
  \caption{(a) The vector field $\mathbf{e}(\bar{\boldsymbol{\theta}})$ is indicated by arrows. The winding number $\mu({\bar{\boldsymbol{\theta}}})$ is a topological index which measures how many times the vector field rotates along $S_{\bar{\boldsymbol{\theta}}}$.
  (b) The total intersection number $I$ is a homotopy invariant of a ZM and counts the minimal number of periodic configurations along that ZM. (c) A ZM with a deformed trajectory has the same total intersection number as (b).
  }
  \label{fig:homo}
\end{figure}

To understand why the two models discussed above behave so differently in presence of non-linearity, we introduce a prescription of defining topological indices in terms of the Poincaré–Hopf index\cite{PoincareHopf}
that accommodates non-linear constraints as well. We will show that this topological index not only recasts the polarization $Q$ defined in the linear model, but can also be extended to an intersection number to identify topologically protected solitons. The definition of the index involves a non-linear map $\mathbf{e}(\boldsymbol{\theta})$ which can be thought of as the vector field on the space of generalized coordinates as shown in Fig. \ref{fig:homo} (a). In the isostatic case ($m=n$), for a solution $\bar{\boldsymbol{\theta}}$ satisfying $\mathbf{e}(\bar{\boldsymbol{\theta}})=\mathbf 0$, we can define an index $\mu(\bar{\boldsymbol{\theta}})$ by computing the winding number of the map $\mathbf{e}(\boldsymbol{\theta})$ on the $(n-1)$-dimensional sphere enclosing $\bar{\boldsymbol{\theta}}$, $S_{\bar{\boldsymbol{\theta}}}$ by integrating the differential form
\begin{equation}
    \label{eq:windingnumber}
    \mu(\bar{\boldsymbol{\theta}}) = \frac{1}{(n-1)! A_{n-1}}\oint_{S_{\bar{\boldsymbol{\theta}}}} \frac{e_{i_1} d e_{i_2}\wedge ... \wedge d e_{i_n} \epsilon^{i_1,i_2,...,i_n}}{(e_1^2+e_2^2+...+e_n^2)^{n/2}},
\end{equation}
where $A_{n-1}$ is the surface area of a unit $(n-1)$-dimensional sphere. When, for example, $n=2$, it yields the so-called first Chern number which frequently appears in classifying the topology in electronic band structures. 

When the Jacobian is full rank, $\mu(\bar{\boldsymbol{\theta}}) = \textrm{sgn}[\textrm{det}( \partial e_{i}(\bar{\boldsymbol{\theta}}) / \partial \theta_j )]$\cite{Fonseca2005}. Under this condition, the configuration $\bar{\boldsymbol{\theta}}$ is structurally stable meaning that $\mu(\bar{\boldsymbol{\theta}})$ is invariant under small, continuous deformations of the constraint functions $\mathbf{e}(\boldsymbol{\theta})$. The idea of topological protection in a linear theory can now be cast as the following: without any symmetry, the phonon spectrum is characterized by a $\mathbb{Z}_2$ invariant protected by a bulk gap that closes when the Jacobian is not full rank.

For the KL chain and the origami chain with a uniform solution $\bar{\boldsymbol{\theta}}$, the topological number $\mu( \bar{\boldsymbol{\theta}} )$ can be simplified to
$\mu(\bar{\boldsymbol{\theta}}) = \textrm{sgn} \{  [\partial_1 f(\bar{\theta},\bar{\theta})]^n - [-\partial_2 f(\bar{\theta},\bar{\theta})]^n \}$
which only depends on the magnitude of $\partial_1 f(\bar{\theta},\bar{\theta})$ and $\partial_2 f(\bar{\theta},\bar{\theta})$. Consequently, $\mu(\bar{\boldsymbol{\theta}})=1$ when $|\partial_1 f(\bar{\theta},\bar{\theta})| > |\partial_2 f(\bar{\theta},\bar{\theta})|$ and $\mu(\bar{\boldsymbol{\theta}})=-1$ when
$|\partial_1 f(\bar{\theta},\bar{\theta})| < |\partial_2 f(\bar{\theta},\bar{\theta})|$.
Therefore, $\mu(\bar{\boldsymbol{\theta}}) = 2 Q - 1$, where $Q$ is the topological polarization discovered by Kane and Lubensky \cite{tm1}.



So far, the above discussion only applies to a specific solution or an (isolated) zero-energy configuration which we now extend to derive an invariant that classifies the topology of the non-linear ZMs. To do so, we look at this topological index from another perspective by first defining a tangent $d$-form 
\begin{equation}\label{eq:tangentform0}
    T^{i_1 \cdots i_d} = \epsilon^{i_1 \cdots i_d j_1 \cdots j_{n-d}} \partial_{j_1} e_1 \cdots \partial_{j_{n-d}} e_{n-d},
\end{equation}
where $d$ denotes the dimension of the non-linear ZM. 
Since $T^{i_1 \cdots i_d}(\boldsymbol{\theta_{i_1}} \cdots \boldsymbol{\theta_{i_d}})=0$ for any vector $\boldsymbol{\theta_{i_j}}$ normal to the space of ZMs, we can think of $T^{i_1 \cdots i_d}$ as defining the tangent space of non-linear ZMs.
For an open KL chain, the number of constraints is one less than the number of the degrees of freedom, and so $d=1$. Then $T$ is a vector field that is everywhere tangent to a non-linear ZM. In this case, the non-linear ZM can be found as the solution to the first-order differential equation
$\partial_s \boldsymbol{\theta}(s) = T\left[ \boldsymbol{\theta}(s) \right]$.
So long as $T(\boldsymbol{\theta})$ is a smooth non-vanishing function of $\boldsymbol{\theta}$, the integral curves of $T(\boldsymbol{\theta})$ will be smooth as well. For any surface not parallel to the tangent $T(\boldsymbol{\theta})$, we can define an intersection number at the point $\bar{\boldsymbol{\theta}}$ where the ZM intersects with the surface as
$\nu(\bar{\boldsymbol{\theta}}) = \textrm{sgn}~\left[ T( \bar{\boldsymbol{\theta}} ) \cdot \hat{N}( \bar{\boldsymbol{\theta}} ) \right]$
where $\hat{N}( \bar{\boldsymbol{\theta}} )$ is the unit normal to the surface at $\bar{\boldsymbol{\theta}}$. Alternatively, we can define a vector $\mathbf{g}(\boldsymbol{\theta})=(e_1,e_2,...,e_{n-1},h)$ 
where $h$ is the function describing the surface. Then $\nu( \bar{\boldsymbol{\theta}} )$ can be computed as
\begin{equation}
    \label{eq:windingnumber2}
    \nu( \bar{\boldsymbol{\theta}} ) = \frac{1}{(n-1)!A_{n-1}} \oint_{S_{\bar{u}}} \frac{ d g_{j_1} \wedge d g_{j_2} \wedge \cdots \wedge d g_{j_n} \epsilon^{j_1 j_2 ... j_n } }{(g_1^2+g_2^2+...+g_n^2)^{n/2}},
\end{equation}
similar to the way $\mu$ was defined earlier in Eq.~\ref{eq:windingnumber}. This results in
$\nu( \bar{\boldsymbol{\theta}} ) = \textrm{sgn}~[\textrm{det}~\nabla g( \bar{\boldsymbol{\theta}} )]$
when the Jacobian of $g$, denoted $\nabla g( \bar{\boldsymbol{\theta}} )$, is full rank. 

The function $h$ can be thought as an auxiliary constraint used to obtain information of a non-linear ZM. For example, if we choose $h=e_n= f(\theta_n,\theta_1)$ as defined previously, then $\nu( \bar{\boldsymbol{\theta}} )$ would be the index $\mu( \bar{\boldsymbol{\theta}} )$ defined in Eq.~\ref{eq:windingnumber}. There are many choices for the surface or the auxiliary constraint, but which one gives us a useful index (intersection number) to classify non-linear ZMs?


To answer this question and further illustrate the idea, in what follows, we will specialize to a single unit cell with a two-dimensional space specified by $(\theta_1, \theta_2)$ and consider the surface (or the line in a two-dimensional space) $h$ specified by $\theta_2 - \theta_1=0$. In this example, every time the non-linear ZM for a single unit cell (UCZM) crosses this plane at $\bar{\boldsymbol{\theta}}$, we can associate an index $\nu(\bar{\boldsymbol{\theta}})$ with the intersection point as defined above [see Fig.~\ref{fig:homo} (b)]. With this in mind, for continuous deformations of the trajectory of the UCZM [see Fig.~\ref{fig:homo} (c)], new uniform configurations can be created or annihilated in pairs of opposite indices, but the total intersection number 
$I=\sum_i{\nu}(\bar{\boldsymbol{\theta}}_i)$
of the UCZM remains invariant.

The idea of topological protection, defined as it is in terms of an inherently linear concept of the phonon spectrum as highlighted before, can be carried over in a robust way to non-linear mechanical systems. To be more precise, we imagine that we have a family of constraints $\mathbf{e}_{\mathbf{t}}(\boldsymbol{\theta})$ depending on various parameters of the mechanical system denoted by ${\mathbf{t}}$. In the case of the KL chain, ${\mathbf{t}}$ would contain, for example, the rotor length $r$ and spring length $L$. The idea of topological protection can be understood in the following way: the space of ZMs for one set of parameters can be continuously deformed into the the space of ZMs of another set of parameters as long as no ZM intersects with others or itself during the process of deformation. Then it will become clearer why the KL chain and the origami chain behave so differently despite their superficial similarity after computing the intersection number of a single unit cell for the KL chain and the origami chain. . 

\begin{figure}
  \includegraphics[width=8cm]{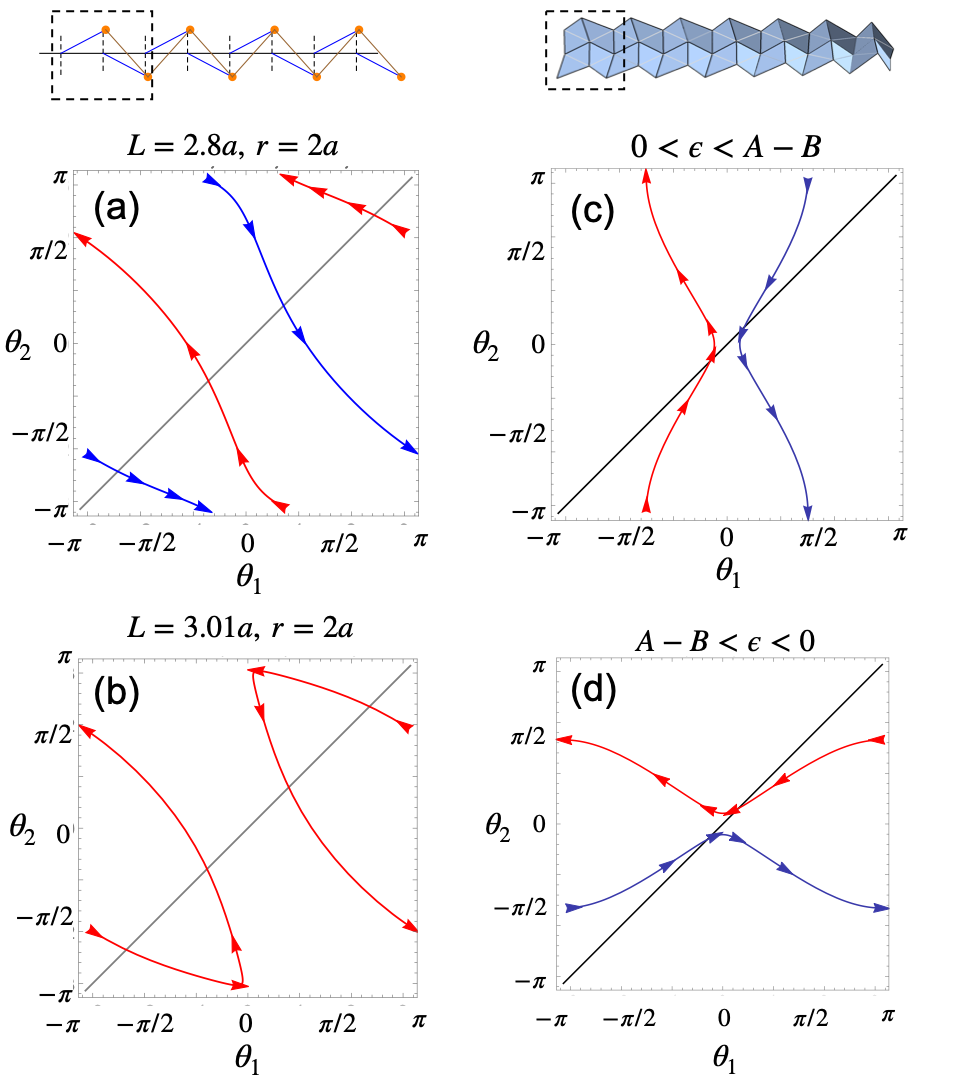}
  \caption{(a)-(b) are the spaces of ZMs of a single unit cell for the KL chain. (c)-(d) are the spaces of ZMs of a single unit cell for the origami chain.}
  \label{fig:chains}
\end{figure}

First, Fig.~\ref{fig:chains} (a)-(b) show the solutions to Eq. (\ref{eq:fourbarconstraint}) for a unit cell of the KL chain (consisting of a pair of rotors). Uniform solutions, namely, $\theta_1=\theta_2$ (there are four) correspond to the points where the non-linear UCZMs cross the plane $\theta_1 - \theta_2=0$. We note that, in the non-linear model, the trajectory of a non-linear UCZM passes through either two or all four of these (uniform) solutions depending on the values of $L$, $r$, and $a$.
The total intersection number $I$ of a non-linear UCZM satisfies the following condition: when $a < L < 2 r - a$, there are two distinct UCZMs with $I=+2$ [blue in Fig.~\ref{fig:chains} (a)] and $I=-2$ [red in Fig.~\ref{fig:chains} (a)]. Thus, each UCZM passes two distinct uniform solutions at least twice and these two uniform solutions are necessarily connected via the trajectory of the UCZM. This case is known as the ``flipper'' phase of the KL chain, characterized by flipper solitons whose existence is topologically protected. When $2 r - a < L < 2 r + a$, on the other hand, we have only one UCZM with a total intersection number $I=0$ [this UCZM passes through all four solutions as in Fig.~\ref{fig:chains} (b)]. This is dubbed the ``spinner'' phase. In this phase, the trajectory of the UCZM can be continuously deformed by tuning, {\it e.g.} $L$, such that all four solutions get annihilated in pairs of opposite intersection numbers exactly at $L=2r+a$, and no solution exists beyond that.

Next, we consider the origami chain. A single unit cell in this model is described by Eq.~\ref{eq:origamiconstraint}.
The uniform solutions are given by the zeros of
$f(\theta,\theta)=(A - B) \sin^2 (\theta/2) + \epsilon$,
which only exist when $(B-A)/\epsilon > 1$. As shown in Fig.~\ref{fig:chains} (c)-(d), there are two distinct regimes: ({\it i}) $0<\epsilon<A-B$, and ({\it ii}) $A-B<\epsilon<0$, both of which have two uniform solutions with opposite sign of $\nu$ and the two UCZMs correspond to the intersection number of $I = +1$ [blue in Fig~\ref{fig:chains} (c) or (d)] or $I = -1$  [red in Fig~\ref{fig:chains} (c) or (d)]. As seen in Fig~\ref{fig:chains} (c)-(d), each UCZM crosses the line defined by $\theta_1 = \theta_2$ at least once. If the system is distorted, it is possible to cross this line multiple times, but the total intersection number remains unchanged. We conclude that the existence of uniform solutions is, indeed, topologically protected. To eliminate them, it is necessary to distort the system through a topological phase transition by joining the trajectories of the two UCZMs. Ultimately, this requires tuning the system through one of the two situations: $\epsilon = 0$ or $A-B + \epsilon = 0$.



It is clear that when a UCZM has a total intersection number $|I| \geq 2$, it must have at least two uniform solutions joined by a smooth trajectory. However, this does not immediately extend to a larger chain of $n$ ($n>2$) unless the following (sufficient) condition $P$ is met: for a given UCZM, either the map from $\theta_i$ to $\theta_{i+1}$ $\forall i$ or the reverse map is injective. 

\begin{figure}
  \includegraphics[width=8cm]{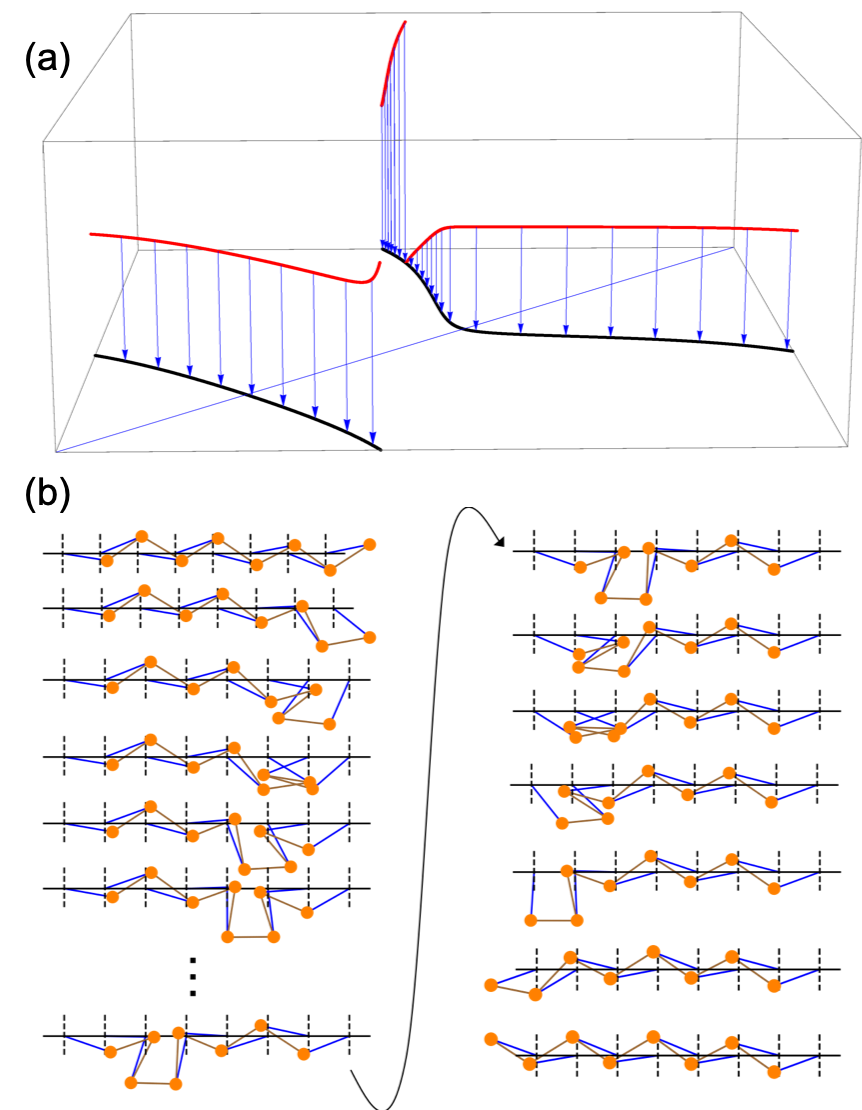}
  \caption{(a) The ZM for the $n=2,3$ KL chain (the flipper case). The black curve $C_2$ on the bottom plane is a single loop on two-dimensional torus, and the red curve $C_3$ is a single loop on three-dimensional torus. (b) A soliton on the disordered KL chain.}
  \label{fig:disorders}
\end{figure}

Lets take the flipper for an example and denote a ZM for the $n$-unit cell chain by $C_n$. In this notation, the black curve on the bottom plane in Fig.~\ref{fig:disorders}(a) is $C_2$ and the red curve is $C_3$. Since, in this case, we have $|I|=2$, the projection of $C_{3}$ onto a constant $\theta_{3}$ plane always yields $C_2$ (it, in fact, extends to $|I|\ge2$). This statement can be understood in the following way: we are looking for a solution for $f(\theta_2,\theta_3)=0$ provided $f(\theta_1,\theta_2)=0$. A sufficient condition for this is that the solution of $f(\theta_2,\theta_3)=0$ on the $\theta_2-\theta_3$ plane wraps around $\theta_2$ at least once
(this holds when $|I|\ge2$) guaranteeing a $\theta_3$ for a given $\theta_2$ that also satisfies $f(\theta_1,\theta_2)=0$.
If the above condition is met, there must exist at least one $\theta_3$ for a given $(\theta_1, \theta_2)$ that satisfies both the constraints. Thus, for each point on the black curve $C_2$, we can always find at least one point on the red curve $C_3$ projected onto it. 

We can now prove that the two uniform solutions are connected by $C_3$ which we have shown to hold for $C_2$ previously. This we prove by contradiction. If we assume that there are two disconnected parts of $C_3$ while $C_2$ is connected, there must exist two points that have the same $\theta_1$ and $\theta_2$ but distinct $\theta_3$. However, this contradicts the fact that the map from $\theta_3$ to $\theta_2$ is injective, and thus, $C_3$ must be connected. The argument can easily be generalized to $C_n$ for $n>3$. Thus, we conclude that there must exist at least two uniform solutions joined by a ZM in a $n$-unit cell chain when a UCZM corresponds to a total intersection number $|I| \geq 2$ and satisfies the condition $P$ mentioned above. This ZM is a soliton (for the non-linear model) that is topologically protected and robust to disorders. We emphasize, a soliton of this kind exists even in a disordered ($a < L_i < 2 r - a$, $L_i$ chosen randomly) KL chain which has the total intersection number $I =\pm 2$ in each unit cell as shown in Fig.\ref{fig:disorders} (b).



We conclude by emphasizing that new topological indices can be generated in similar manners following our prescription to classify non-linear ZMs. For instance, a $n-1$-dimensional sphere around an isolated zero-energy configuration (solution) is chosen in this work as the base manifold to construct a bundle with $\mathbb{Z}$-type topological invariant. For higher-dimensional manifolds of such solutions, different choices of the base manifold can lead to different types of topological invariants \cite{Ktheory}. Exploring the physical significance of those topological indices constitutes a new direction of searching for novel topologically protected non-linear ZMs in the future.

\section*{Acknowledgement}
BC and CDS were partially supported by EFRI 1240441 and CDS was partially supported by DMR 1822638. KR thanks the sponsorship, in part, by the Swedish Research Council.

\bibliographystyle{unsrt}
\bibliography{bib}

\appendix

\newpage

\section{The origami chain}\label{sec:origami}

\begin{figure}
  \includegraphics[width=8cm]{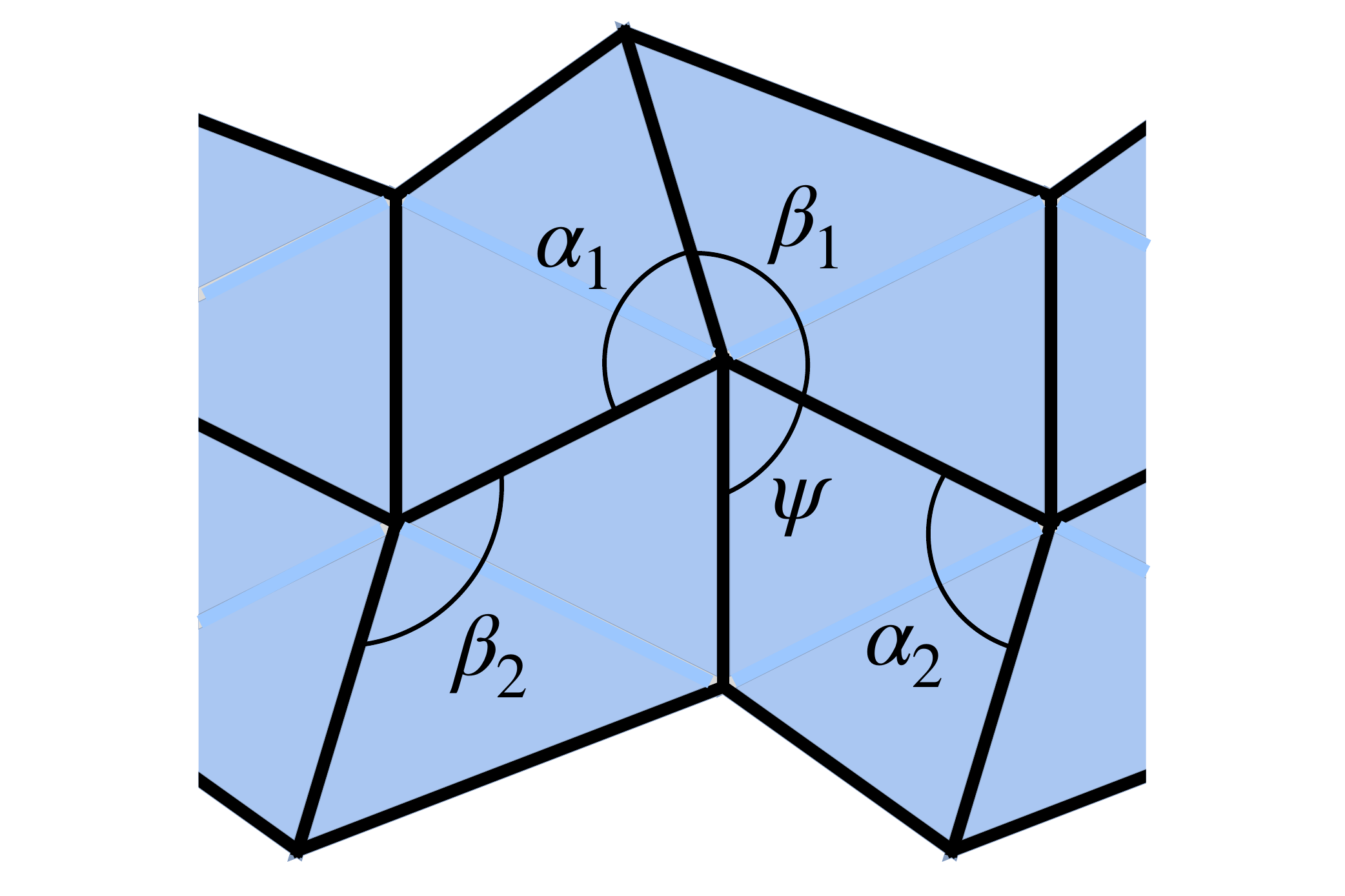}
  \caption{The origami topological chain.}
  \label{fig:schematic}
\end{figure}

The origami chain is a periodic origami fold pattern of degree-4 vertices constructed from quadrilaterals as shown in Fig. \ref{fig:schematic} \cite{tm2}. Each vertex, because it has a one-dimensional configuration space, can be parametrized by the fold angle of a single vertex so any finite number of vertices will have one degree of freedom. There are two degree-4 vertices in each unit cell with interior angles $\{ \alpha_i, \beta_i, \phi_i, \psi_i \}$ for the $i^{th}$ vertex. When the interior angles around each internal vertex add up to $2 \pi$, the fold pattern can be realized as a flat structure; here, we will extend the calculation of Ref. \cite{tm2} to the more general case of arbitrary interior angles.

The law of cosines applied to each vertex gives a constraint on the dihedral angles $\theta$ as,
\begin{eqnarray}
   & &  \cos \left( \alpha_1 \right) \cos \left( \psi_1 \right) + \sin \left( \alpha_1 \right) \sin \left( \psi_1 \right) \cos \theta_1 \\
   & & =  \cos \left( \beta_1 \right) \cos \left( \phi_1 \right) + \sin \left( \beta_1 \right) \sin \left( \phi_1 \right) \cos \theta_2 \nonumber \\
   & & \cos \left( \alpha_2 \right) \cos \left( \psi_2 \right) + \sin \left( \alpha_2 \right) \sin \left( \psi_2 \right) \cos \theta_2 \\
   & & =  \cos \left( \beta_2 \right) \cos \left( \phi_2 \right) + \sin \left( \beta_2 \right) \sin \left( \phi_2 \right) \cos \theta_3 \nonumber
\end{eqnarray}
After some manipulation, we obtain
\begin{eqnarray}
A \sin^2 \left( \frac{\delta \theta_1}{2} \right) - B \sin^2 \left( \frac{\delta \theta_2}{2} \right) + \epsilon = 0,
\end{eqnarray}
where $\delta \theta_i = \pi - \theta_i$ and
\begin{eqnarray}
    A &=& \sin \alpha_1 \sin \alpha_2 \sin \psi_1 \sin \psi_2, \nonumber \\
    B &=& \sin \beta_1 \sin \beta_2 \sin \phi_1 \sin \phi_2, \nonumber \\
    2 \epsilon &=&
        \sin (\alpha_2) \sin (\psi_2) \left[ \cos (\alpha_1+\psi_1)-\cos (\beta_1) \cos (\phi_1) \right] \nonumber \\
    & & + \sin(\beta_1) \sin (\phi_1) \left[ \cos (\alpha_2) \cos (\phi_2)-\cos (\beta_2+\phi_2) \right] \nonumber.
\end{eqnarray}
When the Gaussian curvature of both vertices is zero, $\epsilon = 0$.

\end{document}